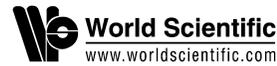

# Photonuclear reaction in $^{67}$Zn

Sylvian Kahane [*,‡] and Raymond Moreh [†,§]

*P. O. Box 1630, 84965 Omer, Israel

†Department of Physics,
Ben-Gurion University,
Beer-Sheva 84120, Israel
‡sylviankahane@gmail.com
§moreh@bgu.ac.il

Mono-energetic $\gamma$-beams ($\Delta \sim 10\,\text{eV}$) based on thermal neutron capture, in a nuclear reactor, using the Mn$(n,\gamma)$ reaction were utilized for generating a fast neutron source from Zinc, via the $^{67}$Zn$(\gamma,n)$ reaction. One of the incident $\gamma$-lines of the Mn source at $E_\gamma = 7244\,\text{keV}$, photoexcites by chance a resonance level in $^{67}$Zn, with subsequent emission of neutrons at an energy of 191 keV. The cross-section for this process was measured and found to be $\sigma(\gamma,n) = 252 \pm 41$ mb with an intensity of the order of $10^4$ n/s. The angular distribution of the 191 keV neutron group was also measured.

## 1. Introduction

We present a method of producing neutron sources based on the $(\gamma, n)$ reactions on various elements using high energy photons of 7–10 MeV originated in the thermal $n$-capture in a nuclear reactor.[1] Therefore, such neutron sources are reactor-driven and require three stringent conditions for their production to be feasible: (1) one of the incident $\gamma$-lines produced by the $(n,\gamma)$ source should resonantly photoexcite, *by chance, an isolated* nuclear level in the target (sample); (2) the photoexcited level should be an *unbound* level higher than the threshold energy for neutron emission; (3) the cross-section for $n$-emission should be high enough for producing a neutron source of relatively high intensity. The obtained intensities are of the order of $10^3$–$10^5$ n/s, comparable to those of the $(\alpha, {}^9\text{Be})$ type or $^{252}$Cf.[2]

---

§Corresponding author.

S. Kahane & R. Moreh

The present $\gamma$ source is generated by thermal neutron capture on $^{55}$Mn via the Mn$(n,\gamma)$ reaction. This reaction emits a $\gamma$-line at 7244 keV which subsequently de-excites to one of the excited states of $^{56}$Mn. The 7244 keV $\gamma$-line happens to photoexcite resonantly an *unbound* level in $^{67}$Zn in which a neutron is emitted via the $^{67}$Zn$(\gamma,n)$ $^{66}$Zn reaction leaving the residual $^{66}$Zn nucleus at its ground state. The emitted neutron has an energy $E_n = 191$ keV with a small energy spread of $\sim 50$ eV caused by kinematics. This neutron source can be created only in an operating nuclear reactor. Such accidental photoexcitation processes are not very rare.[3,4] This may be understood if one considers the large number of the discrete $\gamma$ lines in each photon beam and also the density of nuclear levels occurring in each nucleus at excitations of 7–10 MeV. Both the incident $\gamma$ line and the nuclear level are Doppler broadened having $\Delta \sim 10$ eV where the broadening depends on the energy and the nuclear mass. The small width of the incident $\gamma$ is of paramount importance in the capability to excite a single nuclear level in the continuum of the target, where the level density becomes quite large.

## 2. Experimental Method

The experimental system is shown schematically in Fig. 1 where the $\gamma$ source is produced by the Mn$(n,\gamma)$ reaction. Such $\gamma$ sources are of huge intensities, in our case being produced by using 577 g of metallic Mn mounted on a tangential beam tube, near but outside the core of the Israel Research Reactor 2 (IRR-2). Typical thermal neutron flux, at the $\gamma$-source in this reactor, is of the order $\sim 10^{12}$ n/cm$^2$/s.

The Manganese $(n,\gamma)$ source was in the form of six separated metallic discs each 3 mm thick and 7.5 cm diameter with a spacing of 2 cm from one another. Each Manganese disk was rigidly held in position by graphite ring holders which also served for conducting and dissipating the heat (generated during Reactor operation) to the surrounding structure. The reason is that Mn has a high thermal $n$-capture cross-section of $\sigma_n = 13.4$ b which leads to a larger amount of heat being generated in the sample. The isotope $^{25}Mn^{55}$ is the single stable isotope and the source mass is hence, 577.3 gm. With this mass, the above cross-section and the above neutron flux, a total production $\gamma$ rate of $\sim 8.5 \times 10^{13}$ photons/s is predicted in $4\pi$, not accounting for self-absorption. The distance between the $(n,\gamma)$ source and the target position is $\sim 6$ m. The resulting $\gamma$ beam was collimated and neutron-filtered. Intensities of $\sim 10^6$ photons/cm$^2$/s (for the strong $\gamma$ lines) were measured at the target position.

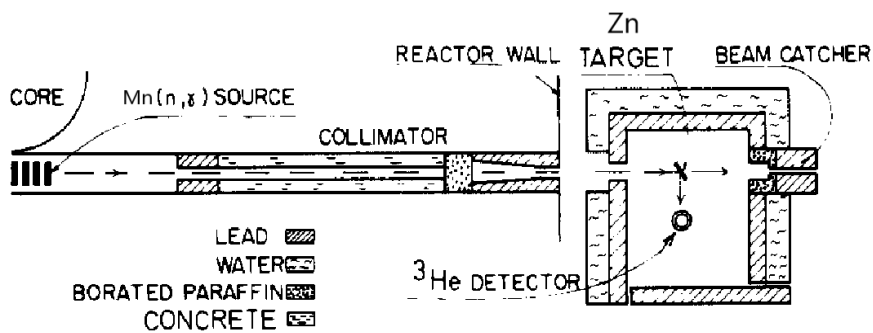

Fig. 1.   Schematic diagram of the experimental setup (not to scale) showing the $(n,\gamma)$ source set inside a beam tube tangential to the reactor core. The high intensity $\gamma$-beam was neutron-filtered using a 40 cm borated parrafin absorber (for reducing the neutron background inside the scattering chamber). The scattering chamber is surrounded by walls of a 15 cm thick lead shielding.

The $^3$He detector, placed at 27 cm from the target, is a commercial neutron spectrometer manufactured by Seforad-Applied Radiation Ltd., Emek Hayarden, Israel, based on the gridded ionization counter of Shalev and Cuttler.[11] Nowadays it is manufactured by Bubble Technologies, Inc. (Ontario, Canada). It is a cylindrical detector, 5 cm diameter, 15 cm active height, 1.2 mm thick (steel), filled at 6 atm with $^3$He, 3 atm with argon and 0.5 atm with methane. Neutron detection relies on the $^3$He$(n,p)$T reaction where $Q = 764$ keV. For *thermal neutrons*, the absorption cross-section is very large, 5300 b, producing a strong peak at 764 keV in the neutron Pulse Height Spectrum (PHS). This peak corresponds to the energy sum of the emitted $p$ and T($^3$H) ions deposited in the ionization chamber. As an illustration, we show in Fig. 2 a PHS taken from a *previous* measurement using practically the same $^3$He detector.[4]

In Fig. 2, only part of the thermal peak is shown and the energy scale of the fast neutrons is selected in such a way that its *zero value* starts at the peak of thermal neutrons. Details of energy calibration are described in Ref. 3.

The signals from the $^3$He $n$-detector were fed through a pre-amplifier to a shaping main amplifier producing Gaussian-shaped signals. The pulses from the amplifier

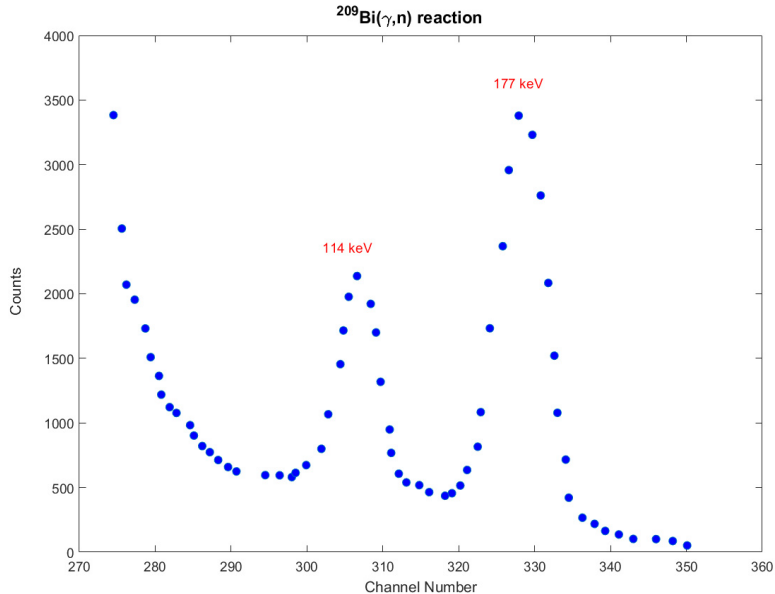

Fig. 2. The 114 keV and 177 keV neutron lines generated using the $^{209}$Bi$(\gamma,n)$ reaction by a $\gamma$-line at 7637 keV obtained from a Cu$(n,\gamma)$ source (taken from Ref. 4).

*S. Kahane & R. Moreh*

were sorted using a Canberra 80 Analyzer where the obtained pulse height spectra could be analyzed for peak area calculation and energy calibration. The best energy resolution was obtained by operating the amplifier with a time constant $\tau \sim 12.8\,\mu$s. It was beneficial to increase $\tau$ with increasing neutron energy to allow for a better charge collection of the $^3$He detector. During the measurements, the observed counting rate was lower than about 3000 cps, mainly due to the good collimation of the incoming gamma beam and to the target detector distance. Hence, no serious pileup effects were observed.

It should be emphasized that the use of a *tangential* beam tube is of paramount importance and one should avoid using a *radial* beam tube for producing the gamma source. This is because in a *radial* beam the amount of gamma and neutron background emerging from the reactor core is so huge that it overwhelms any gamma signal emitted by the $(n,\gamma)$ reactions and also any resonance *n*-signal emitted by the $(\gamma,n)$ reaction by the target. At energies higher than $\sim 5$ MeV, this $\gamma$ background consists mainly of *pileups* and hence cannot produce any resonance photoneutrons from the nuclei of the samples. Note also that such high intensity background photons can be elastically and inelastically scattered from the target and could block the $^3$He detector, being sensitive to $\gamma$-s also. The same is true regarding the neutron background coming from the radial tube.

The direct $\gamma$-beam was passed through a 40 cm borated parrafin absorber (see Fig. 1); located along the tangential beam tube, for reducing the neutron background inside the scattering chamber. A resonance neutron emission via the $(\gamma,n)$ reaction occurs only when one of the strong intensities $\gamma$ lines such as the one at 7244 keV of Fig. 3 happens to overlap by chance and photoexcite an unbound nuclear level in $^{67}$Zn thus emitting 191 keV neutrons as discussed in more detail below.

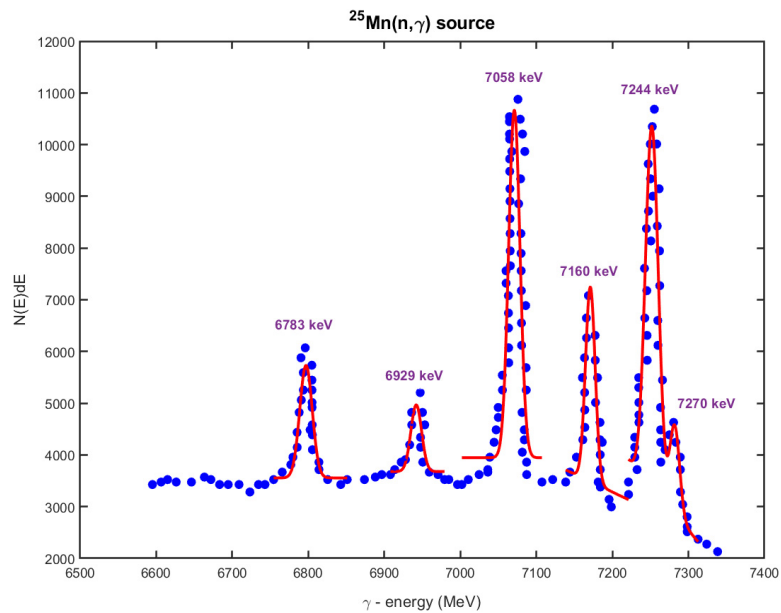

Fig. 3.   The high energy part of the $^{55}$Mn$(n_{\text{th}},\gamma)$ spectrum in the energy range 6600–7300 keV, as measured by Hughes *et al.*[5] (the energy annotations are ours).

## 3. Results

### 3.1. *The* 191 *keV neutron source*

As said, only the 7244 keV $\gamma$-line photoexcites resonantly a nuclear level in $^{67}$Zn. This level is unbound with a neutron separation energy of 7052.5 keV,[6] decaying by neutrons of energy $E_n = 191$ keV, proceeding to the ground state of $^{66}$Zn. The neutron generation process is described in Fig. 4; the detector system has a resolution of 17 keV for thermal neutrons and 24 keV for 1 MeV neutrons. The detector was shielded by wrapping it with 0.5 mm of metallic Cd and 2 mm of metallic Pb.

Natural zinc has five stable isotopes with the following abundances:

| $^{64}$Zn | $^{66}$Zn | $^{67}$Zn | $^{68}$Zn | $^{70}$Zn |
|---|---|---|---|---|
| 48.6% | 27.9% | 4.1% | 18.8% | 0.6% |

The $(\gamma, n)$ threshold for $^{67}$Zn is 7052.5 keV while the Mn$(n, \gamma)$ reaction possesses only four $\gamma$-lines above this energy, namely: 7058, 7160, 7244, 7271-keV, as can be seen in Fig. 3, with the following intensities per 100 neutron captures: 11.4, 6.1, 12.1, 3.1, respectively. This fact means that the emitted neutron source is expected to be relatively clean having a low background.

The intensity of the produced neutron group is influenced by a series of local parameters like the weight and shape of the Mn $\gamma$-source, the Zn sample and the reactor flux. One parameter, of major importance, is the $\sigma(\gamma, n)$ cross-section in the Zn target.

In this work, we are interested to quantify the total cross-section $\sigma(\gamma, n)$ of the resonant photoexcitation process. For this: (1) we determined the differential cross-sections $d\sigma/d\Omega$, at a scattering angle $\theta = 90°$ (see Fig. 1) and (2) we measured the angular distribution of the photoneutrons. These two quantities provided the reported cross-sections. In addition, we also deduced other quantities of physical interest, such as the multipolarity of the emitted neutrons and the nuclear spin of the emitting level.

The $d\sigma/d\Omega$ differential cross-section at $\theta = 90°$, was measured relative to that of 726 keV $n$-group of the $^{49}$Ti$(\gamma, n)$ reaction (measured absolutely in a previous work[8]), by accounting for the relative efficiencies and the differences in the number of nuclei in the samples (otherwise the setups being identical), using the following relation:

$$\left(\frac{d\sigma}{d\Omega}\right)_{E_n} = \frac{Y_\theta^{E_n}}{Y_\theta^{726}} \frac{\mathcal{N}_{726}}{\mathcal{N}_{E_n}} \frac{\epsilon_n^{E_n}}{\epsilon_n^{726}} \left(\frac{d\sigma}{d\Omega}\right)_{726}, \qquad (1)$$

S. Kahane & R. Moreh

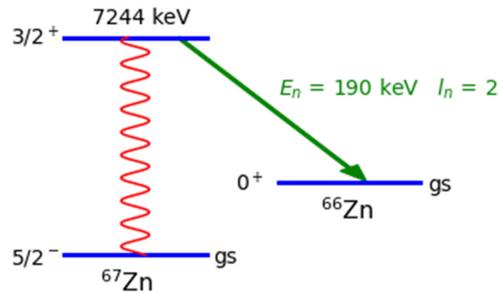

Fig. 4. Photoexcitation of a resonance level in $^{67}$Zn by the 7244 keV $\gamma$-line of the $^{25}$Mn $(n,\gamma)$ reaction and the subsequent neutron emission (with $E_n = 191$ keV and $l_n = 2$) leading to the $J_f^\pi = 0^+$ ground state of $^{66}$Zn.

where $Y$ denotes the measured yields normalized per unit time, at 726 keV and at a neutron energy $E_n$, respectively, $\mathcal{N}$ is the number of nuclei in the two samples (respectively). Franz et al.[7] used an identical detector and monoenergetic neutrons produced with a proton accelerator and the $^7$Li$(p,n)$ $^7$Be reaction, for measuring the relative efficiencies of the detector in the 0.02–2.77 MeV range. We are using their efficiencies in the above formula.

The measured angular distribution is shown in Fig. 5 at 12 angles between 40° and 150°. It was fitted with Legendre polynomial expansion of the form

$$W(\theta) = A_0 + A_2 P_2(\cos\theta), \qquad (2)$$

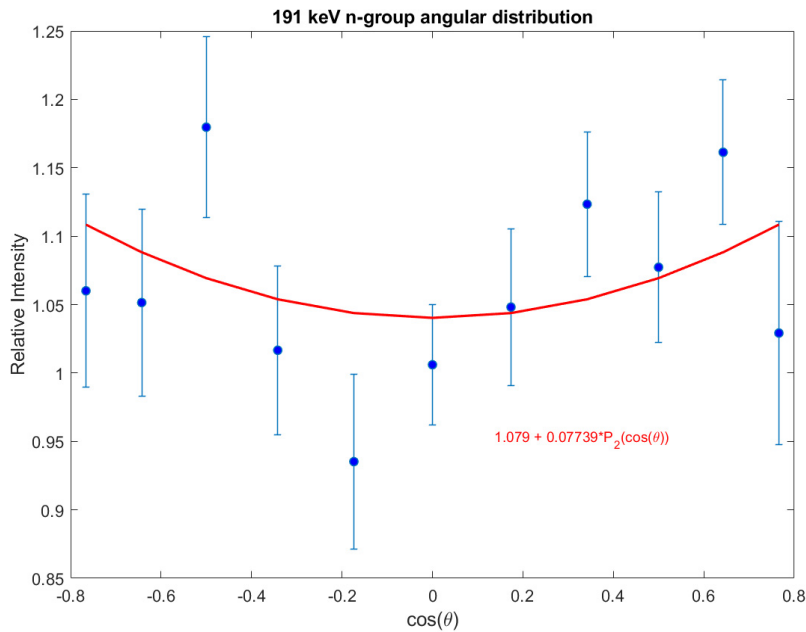

Fig. 5. Angular distribution of the resonance neutron group at 191 keV. The $A2$ parameter has a large error but the ratio $A2/A0 = 0.072$ is small.

which implies a symmetric angular distribution. This distribution was found to be reasonably symmetric around 90° which is characteristic of *n*-emission from a resonance level.[8,9] Since the ground state of $^{67}$Zn has $J^\pi = 5/2^-$, the excited 7244 keV level in $^{67}$Zn seems very likely to have $J^\pi = 3/2^+$, being very strongly excited via $E1$ absorption. This level then emits $l_n = 2$ neutrons proceeding to the $0^+$ ground state of $^{66}$Zn as illustrated in Fig. 4.

Finally, the photoexcitation cross-section for the 7244 keV resonance in $^{67}$Zn was found to be $\sigma(\gamma, n) = 252 \pm 41$ mb.

Due to the recoil of the final nucleus ($^{66}$Zn), the energy of the emitted neutrons varies with the *n*-emission angle relative to the incident $\gamma$-beam, hence we are getting a *n*-group and not a monoenergetic line. These being said, the energy spread is relatively small, of the order of $\sim 50$ eV, hence a quasi-monoenergetic source is obtained.

## 4. Conclusions

We report on a new neutron source with an energy $E_n = 191$ keV using a nuclear reactor and a combination of $(n, \gamma)$ and $(\gamma, n)$ reactions. The method depends on a chance overlap between a discrete $\gamma$-line produced by *n*-capture using the Mn$(n, \gamma)$ reaction and by an *isolated* resonance level in the target nucleus $^{67}$Zn.

The energy of this neutron source at $\sim 191$ keV is of particular interest because it is quasi-monochromatic. It is quite different from the conventional *n*-sources (see a collection of measured PHS in Lorch[10]) such as that of $^{252}$Cf which is spread over a wide energy range of 10 eV to 10 MeV or the $^{241}$AmBe source covering the 100 eV to 10 MeV region.

Due to the complexity of the setup, there is little experience in using this kind of sources. In our group, which pioneered them, use was made mainly in the field of nuclear spectroscopy.[3,4] Their principal advantage is being quasi-monochromatic, and as such, can, in principle, be used for energy and efficiency calibrations of *n*-detectors.

The intensity of this *n*-source can in principle be increased by a factor of $\sim 10$ by using an enriched Zn sample where the abundance of the $^{67}$Zn isotope is increased to e.g., $\sim 41\%$. In such a case, it is expected that the neutron intensity emitted by the $^{67}$Zn$(\gamma, n)$ reaction increases by a factor of $\sim 10$ also.

The intensity of the present *n*-source is comparable with the continuous $\alpha$- and $\gamma$-induced sources but cannot compete with the spallation or accelerator-driven sources.

## ORCID

Sylvian Kahane  https://orcid.org/0000-0003-0985-2819
Raymond Moreh  https://orcid.org/0000-0002-8076-2560

S. Kahane & R. Moreh